\documentclass[aps,pra,amsmath,amssymb,preprintnumbers,twocolumn, showpacs,floatfix ]{revtex4}
\usepackage{amssymb}
\usepackage{color}
\usepackage{graphicx}
\usepackage{subfigure}

\begin{document}
\def\bbm[#1]{\mbox{\boldmath$#1$}}

\title{Amendable Gaussian channels: restoring entanglement via a unitary filter}
\author{A. De Pasquale$^1$, A. Mari$^1$, A. Porzio$^2$, and V. Giovannetti$%
^1 $}
\affiliation{$^1$ NEST, Scuola Normale Superiore and Istituto Nanoscienze-CNR, \\
piazza dei Cavalieri 7, 
I-56126 Pisa, Italy\\
$^2$CNR -- SPIN Complesso Universitario Monte SantAngelo, I-80126 Napoli, Italy}

\begin{abstract}
We show that there exist Gaussian channels which are \textit{amendable}. A
channel is amendable if when applied twice is entanglement breaking while
there exists a \textit{unitary filter} such that, when interposed between
the first and second action of the map, prevents the global transformation
from being entanglement breaking [Phys. Rev. A \textbf{86}, 052302 (2012)].
We find that, depending on the structure of the channel, the unitary filter
can be a squeezing transformation or a phase shift operation. We also
propose two realistic quantum optics experiments where the amendability of
Gaussian channels can be verified by exploiting the fact that it is
sufficient to test the entanglement breaking properties of two mode Gaussian
channels on input states with finite energy (which are not maximally
entangled).
\end{abstract}

\pacs{03.67.Mn, 03.67.Pp, 42.50.Ex}
\maketitle

\section*{Introduction}

\label{sec:conc} Quantum states formally represent the addressable
information content about the system they describe. During their evolution
quantum systems may suffer the presence of noise, for instance due to the
interaction with another system, generally referred as an external
environment. This may cause a loss of information on the system, and leads
to a modification from its initial to its final state. In quantum
communication theory, stochastic channels, that is Completely Positive Trace
Preserving (CPT) mappings, provide a formal description of the noise
affecting the system during its evolution. The most detrimental form of
noise from the point of view of quantum information, is described by the
so-called Entanglement Breaking (EB) maps \cite{EBT}. These maps when acting
on a given system destroy any entanglement that was initially present
between the system itself and an arbitrary external ancilla. Accordingly
they can be simulated as a two--stage process where a first party makes a
measurement on the input state and sends the outcome, via a classical
channel, to a second party who then re-prepares the system of interest in a
previously agreed state~\cite{holevoEBT}.

For continuous variable quantum systems \cite{cv}, like optical or
mechanical modes, there is a particular class of CPT maps which is extremely
important: the class of Gaussian channels \cite{gaussian, review}. Almost
every realistic transmission line (\textit{e.g.}\ optical fibers, free space
communication, \textit{etc.}) can be described as a Gaussian channel. In
this context the notion of EB channels has been introduced and characterized
in Ref.~\cite{HOLEVOEBG}. Gaussian channels, even if they are not
entanglement breaking, usually degrade quantum coherence and tend to
decrease the initial entanglement of the state \cite{buono}. One may try to
apply error correction procedures based on Gaussian encoding and decoding
operations acting respectively on the input and output states of the map
plus possibly some ancillary systems. This however has been shown to be
useless \cite{nogo}, in the sense that Gaussian procedures cannot augment
the entanglement transmitted through the channel (no-go theorem for Gaussian
Quantum Error Correction). Here we point out that such lack of effectiveness
doesn't apply when we allow Gaussian recovering operations to act \textit{%
between} two successive applications of the same map on the system.
Specifically our approach is based on the notion of \emph{amendable channels}
introduced in \cite{mucritico}, whose definition derives from the
generalization of the class of EB maps (Gaussian and not) to the class of EB
channels of order $n$. The latter are maps $\Phi$ which, even if not
necessarily EB, become EB after $n$ iterative applications on the system (in
other words, indicating with ``$\circ$" the composition of super-operator, $%
\Phi$ is said to be EB of order $n$ if $\Phi^n := \Phi\circ \Phi \circ
\cdots \circ \Phi$ is EB while $\Phi^{n-1}$ is not). We therefore say that a
map is amendable if it is EB of order 2, and there exists a second channel
(called \emph{filtering} map) such that when interposed between the two
actions of the initial map, prevents the global one to be EB. In this
context we show that there exist Gaussian EB channels of order $2$ which are
amendable through the action of a proper Gaussian unitary filter (i.e. whose
detrimental action can be stopped by performing an intermediate, recovering
Gaussian transformation).

The paper is structured as follows. In Section I we focus on the formalism
of Gaussian channels, the characterization of EB Gaussian channels and their
main properties. In Section II we explicitly define two types of channels
which are amendable via a squeezing operation and a phase shifter
respectively. For each channel we also propose a simple experiment based on
finite quantum resources and feasible within current technology.

\section{Entanglement breaking Gaussian channels}

\label{sec:gauss}
Let us briefly set some standard notation. A state $\rho$ of a bosonic system with $f$ degrees of freedom is Gaussian
if its characteristic function $\phi_{\rho}(z)=\mathrm{Tr} [\rho W(z)]$ has
a Gaussian form \cite{gaussian}, 
\begin{equation}
\phi_{\rho}(\vec z)=e^{i {\langle\vec{R}\rangle}_{\rho}^ \top \vec{z}-\frac{1%
}{2} \vec{z}^\top {\ \mathbf{V}_{\rho}} \vec{z}}\,.
\end{equation}
 $W(\vec z)$ is the unitary Weyl operator defined on the real vector
space $\mathbb{R}^{2f}$, $W(\vec z):=\exp[i \vec{R}\cdot \Delta \vec{z}]$, where 
\begin{equation}
\Delta=\bigoplus_{i=1}^f 
\left(\begin{array}{cc}0 & 1 \\-1 & 0\end{array}\right)
\end{equation}
is the symplectic form, 
$\vec{R}=\{ Q_1, P_1, \ldots ,Q_f, P_f\}$ and $Q_i$, $P_i$ are the canonical
observables for the bosonic system. $\langle\vec{R}\rangle_{\rho}$ is vector
of the expectation values of $\vec{R}$, and $\mathbf{V}_\rho$ is the
covariance matrix 
\begin{equation}
[\mathbf{V}_\rho]_{ij}=\frac{\langle R_i R_j+R_j R_i\rangle_\rho}{2}-\langle
R_i\rangle_\rho\langle R_j \rangle_\rho\,.
\end{equation}
A CPT map $\Phi$ is called Gaussian if it preserves the Gaussian character
of the states, and can be conveniently described by the triplet $(K,l,\beta)$%
, $l \in \mathbb{R}^{2f}$ and $K$, $\beta$ being $2f \times 2f$ matrices,
which fulfill the condition 
\begin{equation}  \label{eq:CPT}
\beta \geq \pm i [\Delta - K ^\top \Delta K]/2,\quad
\end{equation}
and act on ${\langle\vec{R}\rangle}_{\rho}$ and $\mathbf{V}_{\rho}$ as 
\begin{eqnarray}
\mathbf{V}_{\rho} &\to& {\mathbf{V}_{\Phi[\rho]}}= K^\top \mathbf{V}_{\rho}
K +\beta\, \\
{\langle\vec{R}\rangle}_{\rho} &\to& {\langle\vec{R}\rangle}_{\Phi[\rho]}=
K^\top {\langle\vec{R}\rangle}_\rho + l.
\end{eqnarray}

A special subset of Gaussian channels is constituted by the unitary Gaussian
transformations, characterized by having $\beta =0$: they include multi-mode
squeezing, phase shifts, displacement transformations and products among
them.

The composition of two Gaussian maps, $\Phi=\Phi_2 \circ \Phi_1$, described
by $(K_1,l_1,\beta_1)$ and $(K_2,l_2,\beta_2)$ respectively, is still a
Gaussian map whose parameters are given by 
\begin{eqnarray}  \label{eq:composition}
\Phi_{2}\circ\Phi_1 \longrightarrow \left\{%
\begin{array}{l}
K=K_1 K_2 \\ 
l=K_2^\top l_1 + l_2 \\ 
\beta=K_2^\top \beta_1 K_2+\beta_2.%
\end{array}%
\right.
\end{eqnarray}
Finally, a Gaussian map $\Phi$ is entanglement-breaking~\cite{HOLEVOEBG} if
and only if its matrix $\beta$ can be expressed as 
\begin{equation}  \label{eq:EBTgauss}
\beta=\alpha+\nu,
\end{equation}
with 
\begin{equation}  \label{eq:EBTgauss2}
\alpha \geq \frac{i}{2}\Delta, \quad \mbox{and} \quad \nu\geq \frac{i}{2}%
K^\top\Delta K.
\end{equation}

\subsection{One-mode attenuation channels}

\begin{figure}[t]
\begin{center}
\includegraphics[width= 0.9 \columnwidth]{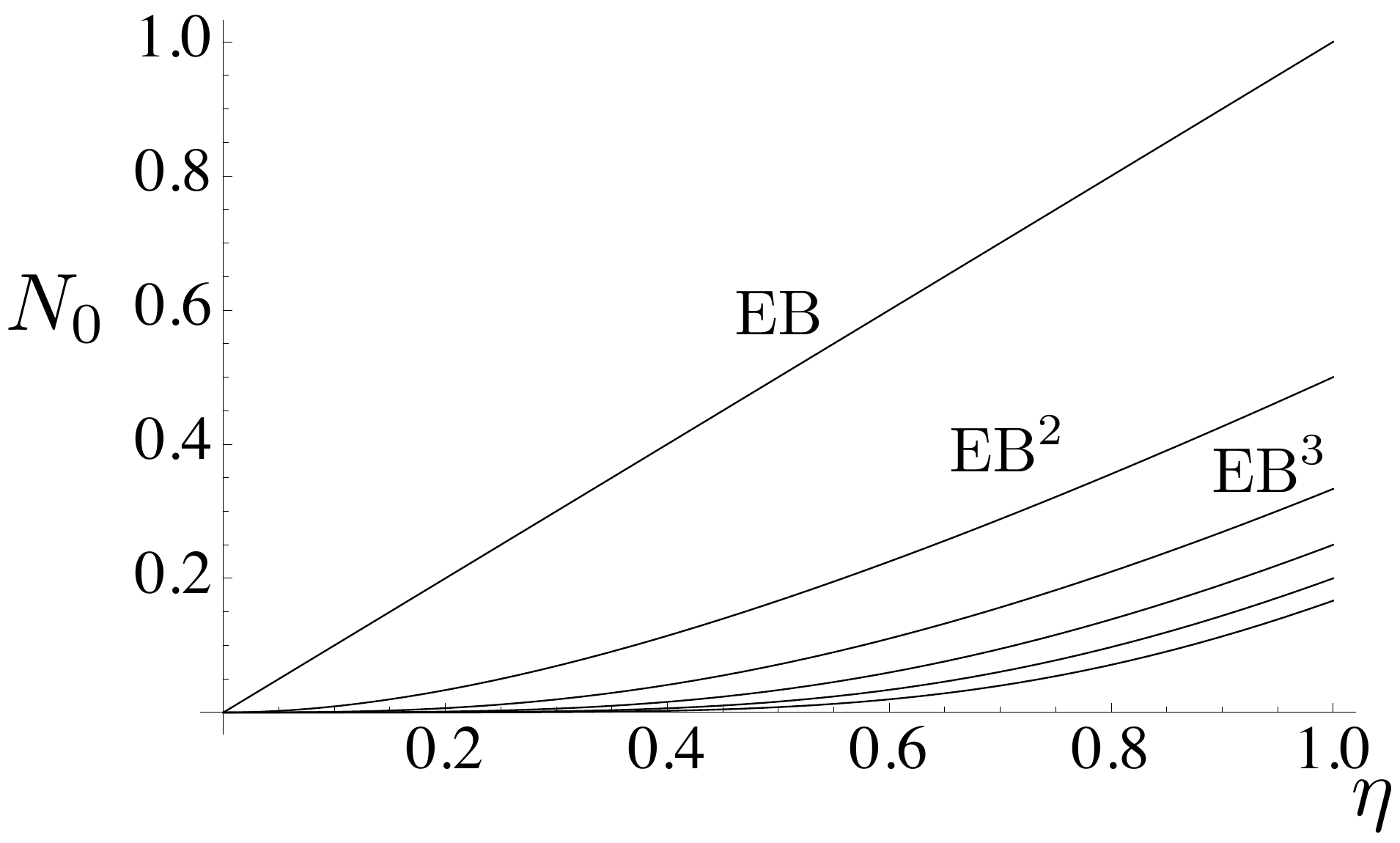}
\end{center}
\caption{Lower boundary of the regions such that $\Phi_{{%
\mathrm{At}}}\in \mathrm{EB}^n$, in the parameter space $\{\protect\eta%
,N_0\} $.}
\label{fig:att}
\end{figure}

One-mode attenuation channels $\Phi_{\mathrm{At}}(N_0,\eta)$ are special
examples of Gaussian mappings such that: 
\begin{eqnarray}
K_{\mathrm{At}}&=&\sqrt{\eta}\; \openone \\
l_{\mathrm{At}}&=&0 \\
\beta_{\mathrm{At}}&=& \left( N_0+\frac{1-\eta}{2}\right)\openone
\end{eqnarray}
where $\openone=\left(
\begin{array}{cc}
1 & 0 \\ 
0 & 1
\end{array}
\right)$, $0\leq\eta\leq1$ and $N_0\geq0$. This transformation can be
described in terms of a coupling between the system and a thermal Bosonic
bath with mean photon number $N=N_0/(1-\eta)$, mediated by a beam splitter
of transmissivity $\eta$. In Ref.~\cite{mucritico} the EB properties of the
maps $\Phi_{\mathrm{At}}(N_0,\eta)$ under channel iteration were studied as
a function of the parameters $\eta^2$ and $N_0$. For completeness we report
these findings in Fig. \ref{fig:att}. 
In the plot the solid lines represent the lower boundaries between the
regions which identify the set of transformations $\Phi_{\mathrm{At}
}(N_0,\eta)$ which are EB of order $n$. They are analytically identified by
the inequalities 
\begin{equation}
N_0 \geq \frac{\eta^{n}}{\sum_{j=0}^{n-1}\eta^{j}}\, ,
\end{equation}
or, in terms of the parameter $N$ which gauges the bath average photon
number, by 
\begin{equation}
N \geq (1-\eta)\frac{\eta^{n}}{\sum_{j=0}^{n-1}\eta^{j}}\, .
\end{equation}
Notice that for $N=0$, $\Phi_{{\mathrm{At}}}^n \notin {\mathrm{EB}} $ for
all finite $n$, that is if the system is coupled with the vacuum (zero
photons) the reiterative application of the map, represented by the action
of a beam-splitter on the input signal, does not destroy the entanglement
between the system and any other ancilla with which it is maximally
entangled before the action of the map.

\subsection{Certifying that a channel is entanglement breaking with non
ideal resources.}

\label{sec:Choi}

It is a well known fact that a map $\Phi$ is EB if and only if when applied
to one side of a maximally entangled state it produces a separable state~%
\cite{EBT}. This fact gives an operationally well defined experimental
procedure for characterizing the EB property of a channel $\Phi$ based on
the ability of preparing a maximally entangled state to be used as probing
state for the map. Unfortunately however, while feasible for finite
dimensional systems, in a continuous variable setting this approach is
clearly problematic due to the physical impossibility of preparing such an
ideal probe state since it would require an infinite amount of energy. Quite
surprisingly, the following property will avoid this experimental issue. 
\newline

{\noindent \textbf{Property (equivalent test states).}} \textit{
Given $\{ |i\rangle; i =1,\cdots, d\}$ an orthonormal set, let $%
\omega=\sum_{i,i'=1}^{d} | i\, i \rangle \langle i'\, i' |$ be an
un-normalized maximally entangled state and $\sigma$  a full-rank $%
d\times d$ density matrix. Then the (normalized) state 
\begin{equation}
\tilde \omega=(\sigma^{1/2} \otimes \openone)\omega (\sigma^{1/2} \otimes %
\openone)  \label{property1}
\end{equation}
is a valid resource equivalent to $\omega$ in the sense that a channel $\Phi$
is EB if and only if $(\openone \otimes \Phi) (\tilde \omega)$ is separable. 
} \newline

\noindent \textit{Proof.} We already know that $\Phi$ is EB if and only if $%
f=(\openone \otimes \Phi) (\omega)$ is separable \cite{EBT}. We need to show
that $f$ is separable if and only if $\tilde f=(\openone \otimes \Phi)
(\tilde \omega)$ is separable. This must be true because the two states
differ only by a local CP map which cannot produce entanglement namely: $%
\tilde f=(\sigma^{1/2} \otimes \openone) f (\sigma^{1/2} \otimes \openone)$
and $f=(\sigma^{-1/2} \otimes \openone) \tilde f (\sigma^{-1/2} \otimes %
\openone)$. \newline

The same property can be extended to continuous variable systems where 
$\omega$ is not normalizable but it can still be consistently interpreted as
a distribution \cite{holevochoi}. Now, let us consider a two-mode squeezed vacuum (TMSV)
state with finite squeezing parameter
$r'$, i.e. 
\begin{equation}
\tilde \omega= \frac{1}{(\cosh r')^2}\sum_{i,i'=0}^{\infty} (\tanh r^{\prime}) ^{i+i'} |i\rangle_{1}
\langle i'| \otimes |i\rangle_{2} \langle i'|,
\end{equation}
 where $\{ |i\rangle; i =1, 2, \cdots \}$ is now the Fock basis.
It can be expressed in the form of Eq.\ (\ref{property1}) by
choosing 
\begin{equation}
\sigma= \text{tr}_2 \{\tilde \omega\}=\frac{1}{(\cosh r')^2}\sum_{i=0}^{\infty} (\tanh r^{\prime})^{2i}
|i\rangle_1 \langle i|
\end{equation}
and therefore the state $\tilde \omega$ is a valid resource for the EB test.  
The previous property implies that \textit{it is
sufficient to test the action of a channel on a two-mode squeezed state with
arbitrary finite entanglement in order to verify if the channel is EB or not.} 
Surprisingly, even a tiny amount of entanglement is in principle enough for the test. However,
because of experimental detection noise and imperfections, a larger value of $r'$ may be 
preferable as it allows for a clean-cut discrimination.

The previous results are obviously extremely important from an experimental point of view
since, for single mode Gaussian channels, one can apply the following
operational procedure:

\begin{itemize}
\item Prepare a realistic two-mode squeezed vacuum state $\tilde \omega$
with a finite value of $r^{\prime }$,

\item Apply the channel $\Phi$ to one mode of the entangled state resulting
in $\tilde f=(\openone \otimes \Phi) (\tilde \omega)$,

\item Check if the state $\tilde f$ is entangled or not.
\end{itemize}

Probably the experimentally most direct way of witnessing
the entanglement of $\tilde{f}$ is to apply the so-called product criterion 
\cite{prodcrit}. In this case, entanglement is detected whenever 
\begin{equation}
\mathcal{W}=\left\langle Q^{2}\right\rangle \left\langle P^{2}\right\rangle <%
\frac{1}{4}  \label{crit}
\end{equation}%
with 
\begin{equation}
Q=\frac{Q_{1}+Q_{2}}{\sqrt{2}},\qquad P=\frac{P_{1}-P_{2}}{\sqrt{2}}\,.
\label{eq:wgauss}
\end{equation}%
We indicate with $Q_{i}$ and $P_{i}$, $i=1,2$, the position and momentum
quadratures associated to each mode of the twin beam. If inequality (\ref%
{crit}) is satisfied, $\tilde{f}$ is entangled and so $\Phi $ is not EB.
This test, is a witness but it does not provide a conclusive separability
proof.  For this reason it is useful to compare it with a
necessary and sufficient criterion. We will use  
the logarithmic negativity $E_{\mathcal{N}}$, which is an entanglement
measure quantifying the violation of the $PPT$ separability criterion \cite%
{PPT}. 
Let $\mathbf{V}_{\tilde{\omega}}$ be the covariance matrix of $\tilde{\omega}
$ written in the block form 
\begin{equation}
\mathbf{V}_\rho=\begin{pmatrix} \mathbf{A} & \mathbf{C}\\
\mathbf{C^\top}&\mathbf{B}\end{pmatrix}\,.
\end{equation}
The entanglement negativity $E_{\mathcal{N}}$ is a function of the four
invariants under local symplectic transformations $\det [\mathbf{A}],\det [%
\mathbf{B}],\det [\mathbf{C}],\det [\mathbf{V}_{\rho }]$ and can be
analytically computed \cite{gaussian}: 
\begin{eqnarray}
E_{\mathcal{N}} &=&\max \{-\ln (2\nu ),0\} \\
\nu  &=&\sqrt{\frac{\Sigma -\sqrt{\Sigma ^{2}-4\det [\mathbf{V}_{\rho }]}}{2}%
}  \label{eq:nu}
\end{eqnarray}%
where $\Sigma =\det [\mathbf{A}]+\det [\mathbf{B}]-2\det [\mathbf{C}]$.
Notice that $\nu $ is the minimum symplectic eigenvalue of the partially
transposed state and can be interpreted as an \textit{optimal product
creterion} since we have that $\tilde{f}$ is entangled if and only if 
\begin{equation}
\nu ^{2}<\frac{1}{4}\,,  \label{eq:nuquadromis}
\end{equation}%
while  Eq.\ (\ref{crit}) is only a sufficient
condition.

Both tests  Eq.\ (\ref{crit}) and (\ref{eq:nuquadromis})
will be used for assessing, in the next section,
the entanglement breaking property
of two possible realization of amendable Gaussian channels. We note that,
direct simultaneous measurements, in a dual-homodyne set-up, on the
entangled sub-systems allow a direct evaluation of the product criterion
\cite{dualhomo}.
While, the experimental evaluation of $E_{\mathcal{N}}$ requires the
reconstruction of the bipartite system covariance matrix that in many cases
can be gained by a single homodyne~\cite{singlehomo}. 

\section{Amendable gaussian maps}

In this section we aim to prove the existence of amendable Gaussian maps
constructing explicit examples and propose experimental setups that would
allow one to implement and test them. To do so we will look for Gaussian
single mode maps $\mathcal{U}$ and $\Phi$, where $\mathcal{U}$ is unitary,
such that 
\begin{eqnarray}  \label{eq:PhiUPhiEB}
\Phi \circ \mathcal{U} \circ \Phi &\in& {\mathrm{EB}} \;, \\
\Phi^2 &\notin& {\mathrm{EB}} \;,  \label{eq:PhiUPhiEB1}
\end{eqnarray}
(notice that the second condition requires that $\Phi$ cannot
be EB). Under these assumptions, it follows that the channel $\Phi^{\mathcal{U%
}}=\mathcal{U}\circ \Phi$ is an EB map of order 2 which can be amended by
the unitary filter $\mathcal{U}^{\dagger}$. Indeed exploiting the fact that local unitary transformation cannot alter
the entanglement, the above expressions imply: 
\begin{eqnarray}
\Phi^{\mathcal{U}}\circ\Phi^{\mathcal{U} }= \mathcal{U} \circ \Phi \circ 
\mathcal{U} \circ \Phi &\in& {\mathrm{EB}} \;,  \label{eq:phiuphiu} \\
\Phi^{\mathcal{U}} \circ \mathcal{U}^{\dagger} \circ \Phi^{\mathcal{U}} = 
\mathcal{U} \circ \Phi^2 &\notin& {\mathrm{EB}}\;.  \label{argum}
\end{eqnarray}
Even though (\ref{eq:PhiUPhiEB}), (\ref{eq:PhiUPhiEB1}) and (\ref{eq:phiuphiu}), (\ref{argum})  are
formally equivalent it turns out that the former relations are easier to be implemented experimentally. For this reason in the following  we will
focus on such scenario.

\subsection{Example 1: Beam splitter-squeezing-beam splitter}\label{EX1} 

\label{example1} 
\begin{figure}[t]
\begin{center}
\includegraphics[width=1\columnwidth]{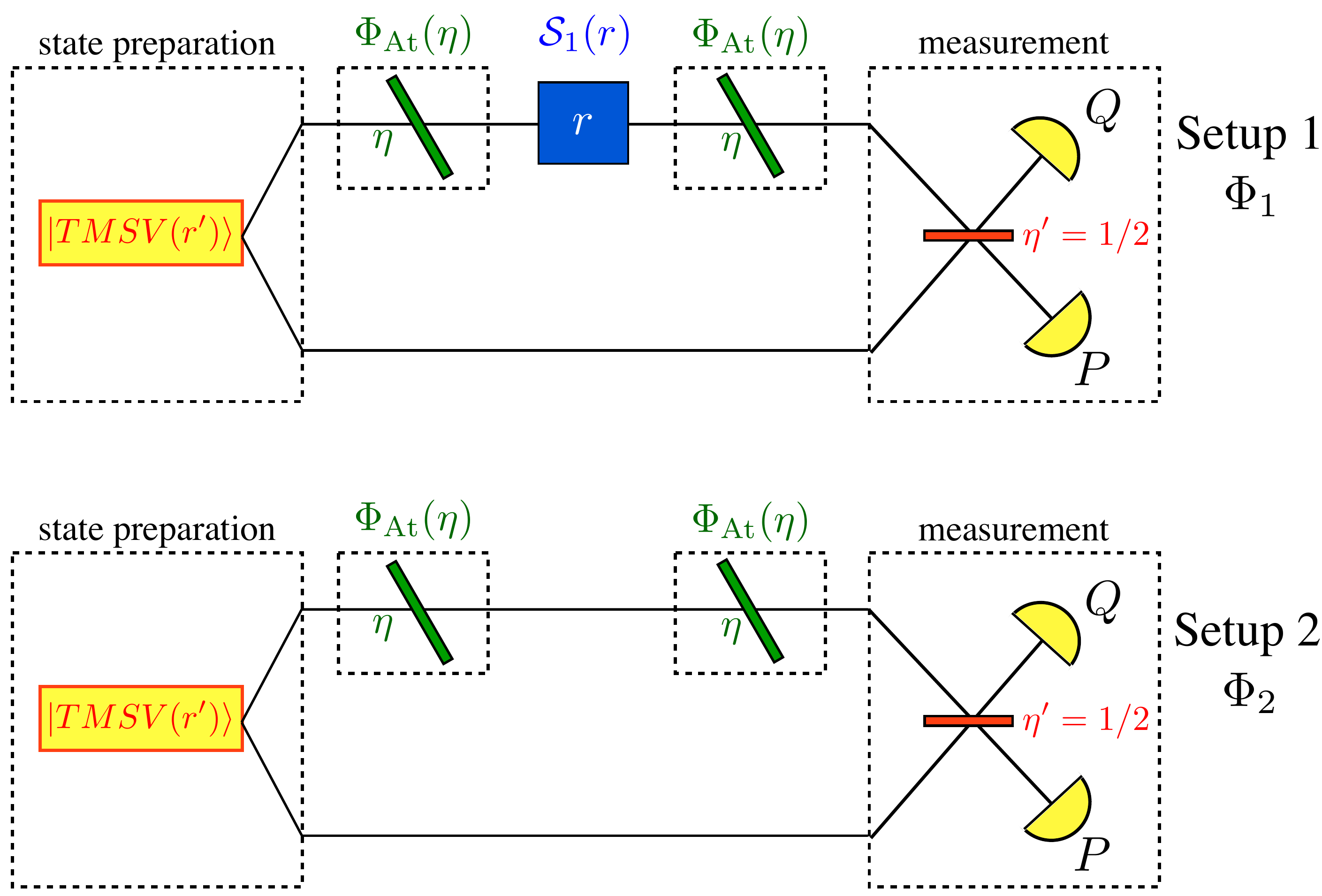}
\end{center}
\caption{(Color online) Schematic of the experimental proposal discussed in Sec.~\ref{EX1}.
 Both setups are divided in three stages: a $|TMSV\rangle$ state is prepared, the desired
   sequence of channels is applied to one mode of the entangled probe, and finally the output state is measured. 
The beam-splitters implement the 
attenuation channels $\Phi_{At}(\protect\eta)$ of Eqs.~(\ref{conc2}), (\ref{conc1}) 
which represent the
transformations $\Phi$ of Eqs.~(\ref{eq:PhiUPhiEB}), (\ref{eq:PhiUPhiEB1}), while the squeezing transformation $%
\mathcal{S}_1(r)$ implements the unitary $\mathcal{U}$. }
\label{fig:simplifiedsetups}
\end{figure}

Here we provide our first example of a channel $\Phi$ and
of a unitary transformation $\mathcal{U}$ fulfilling Eqs.~(\ref{eq:PhiUPhiEB}%
) and (\ref{eq:PhiUPhiEB1}). We will consider two mode
Gaussian maps. By exploiting the property explained in Sec.~\ref{sec:Choi}
regarding the equivalence of test states, without loss of generality we will
apply our channels to twin-beam states with finite squeezing parameter, that
is with finite energy, rather then to  maximally entangled states which
would require an infinite amount of energy to be realized. Eqs.~(\ref%
{eq:PhiUPhiEB}) and (\ref{eq:PhiUPhiEB1}) will be implemented by the two
setups of Fig.~\ref{fig:simplifiedsetups}:

\begin{itemize}
\item The first one (setup $1$) is used to realize the transformation $\Phi
\circ \mathcal{U} \circ \Phi$. It consists in an optical squeezer,
implementing the unitary $\mathcal{U}$, coupled on both sides with a
beam-splitter (one for each side) of transmissivity $\eta$.

\item The second setup (setup $2$ of Fig.~\ref{fig:simplifiedsetups})
instead is used to realize the transformation $\Phi\circ \Phi$: it is
obtained from the first by removing the squeezer between the beam splitters.
\end{itemize}

 As anticipated we will use $|TMSV\rangle$ states as entangled
probes. The aim of the section is to show that by properly choosing the
system parameters, the squeezing and the beam-splitter transmissivities, it
is possible to realize an amendable Gausssian channel.

The transformation induced by the beam splitter can be described by an
attenuation map with $N_0=0$, $\Phi_{\mathrm{BS}_1}(\eta):=\Phi_{\mathrm{At}%
}(0,\eta) $. On the other hand, we indicate as $\mathcal{S}_1(r)$ the
unitary map depending on the real parameter $r$, referring to the action of
an optical squeezer 
\begin{eqnarray}
K_{\mathcal{S}_1}(r)&=&\begin{pmatrix}e^{r}&0\\0&e^{-r}\end{pmatrix} \\
l_{\mathcal{S}_1}&=&0 \\
\beta_{\mathcal{S}_1}&=&0\,.
\end{eqnarray}

\begin{figure}[tbp]
\centering
\subfigure[] {\label{valtilde-a}
\includegraphics[width=0.9 \columnwidth]{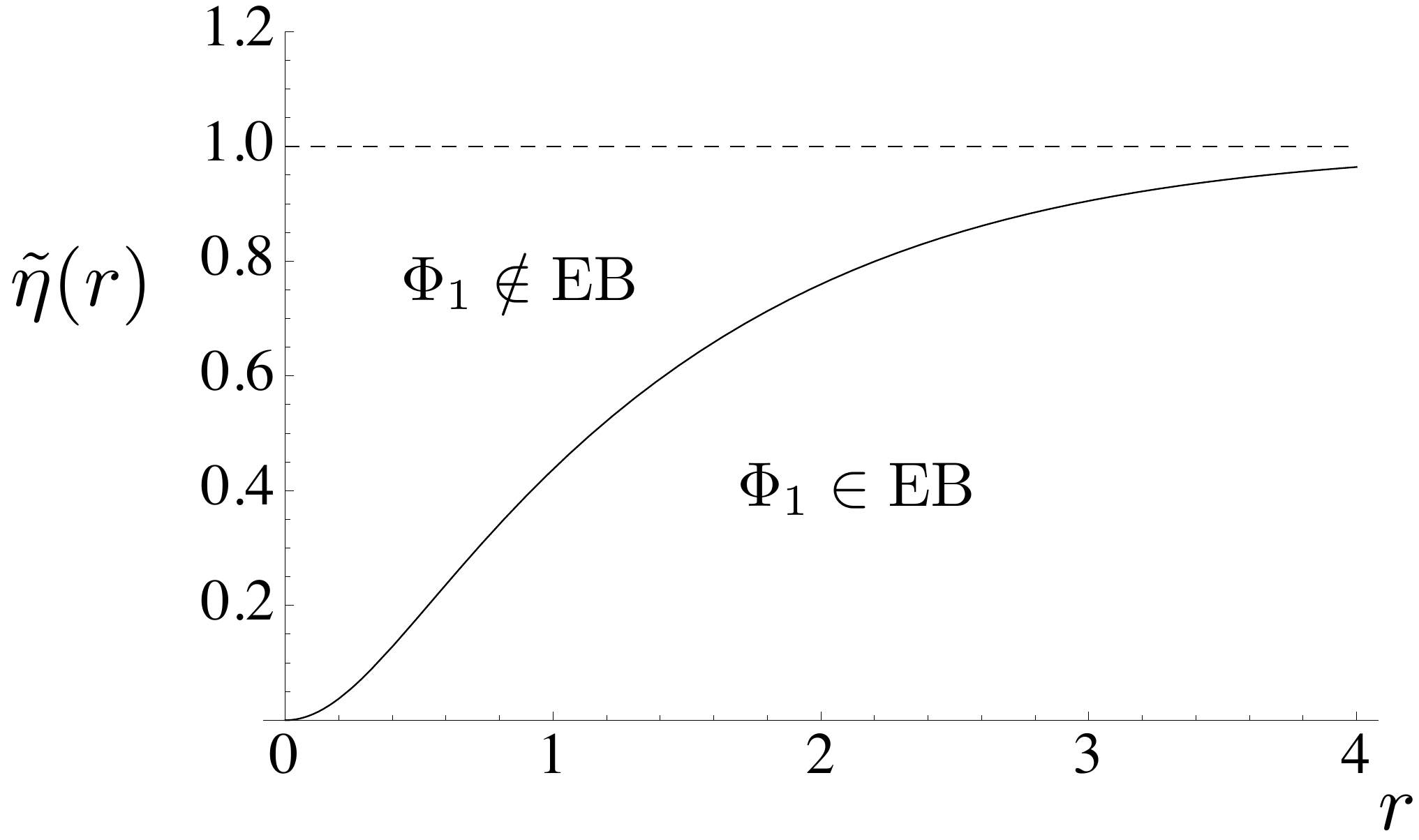}} 
\hspace{5mm} \subfigure[] {\label{}\includegraphics[width=0.9%
\columnwidth]{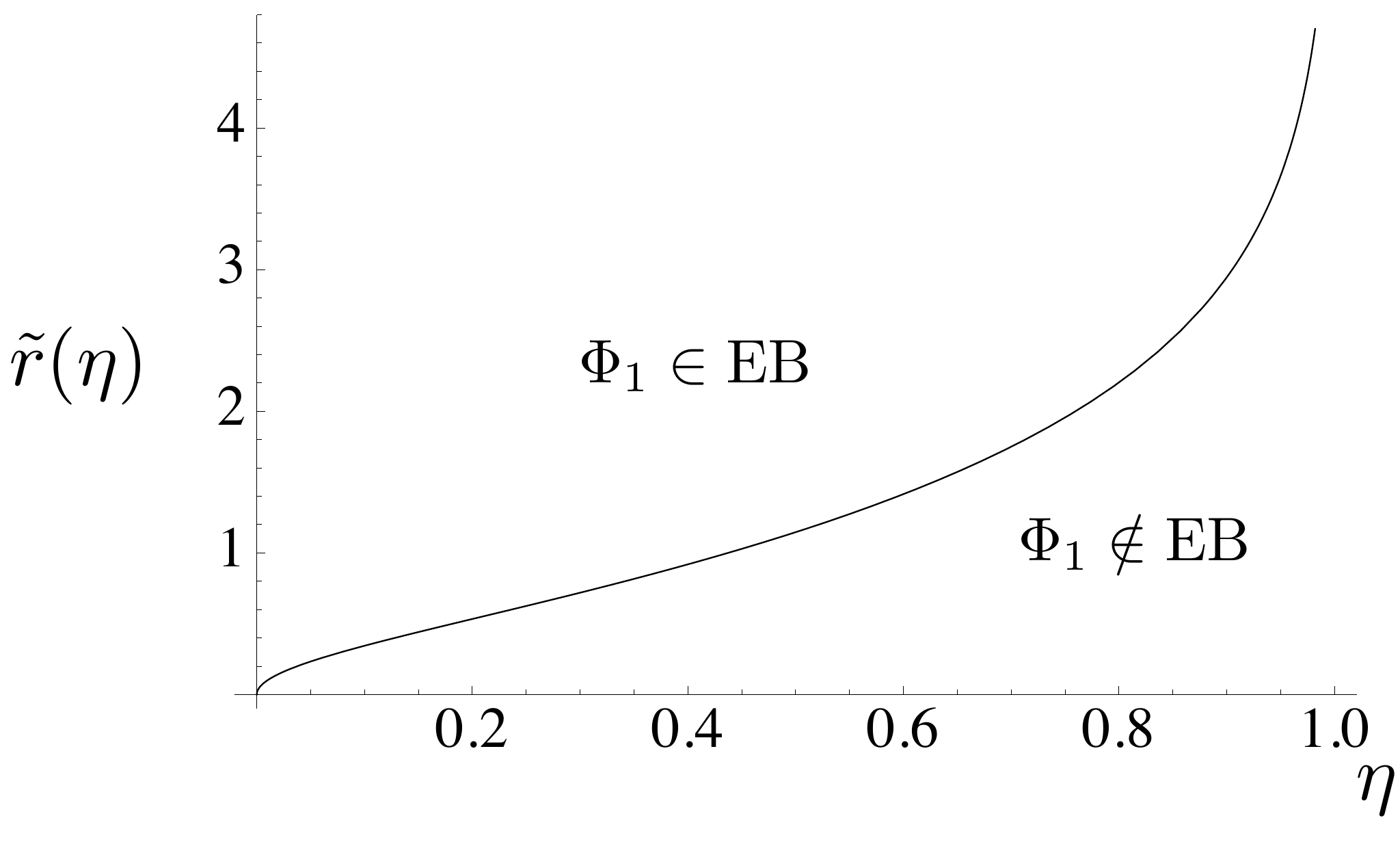}}
\caption{ Lower (a) /Upper (b) bound of the ${\mathrm{EB}} $
region for $\Phi_1$. Notice that in (a) $r$ diverges in the limit of
transmissivity $1$ for the beam splitter, and in the complementary plot (b)
the transmissivity reaches $1$ asymptotically for $r\to \infty$.}
\label{fig:valtilde}
\end{figure}

We set the initial state of the two modes to be a twin-beam $\rho_0
(r^{\prime })=|TMSV \,(r^{\prime })\,\rangle \langle TMSV \,(r^{\prime })\,|$%
, with covariance matrix given by 
\begin{equation}  \label{eq:TMSVcov}
\mathbf{V}_{2s}(r^{\prime })=\frac{1}{2} \begin{pmatrix}\cosh {r'} \openone
& \sinh r' \sigma_z \\ \sinh r' \sigma_z & \cosh r' \openone \end{pmatrix}\,.
\end{equation}
The states at the output of our two setups are described by the following
2-mode density matrices, 
$\rho_{\Phi_1}:= (\Phi_{1}\otimes I)[\rho_{0}] $ and $\rho_{\Phi_2}:=
(\Phi_{2}\otimes I)[\rho_{0}]$ 
 with 
\begin{eqnarray}
&\Phi_1:= \Phi_{1}(\eta,r)=\Phi_{\mathrm{At}}(\eta) \circ \mathcal{S}%
_1(r)\circ \Phi_{\mathrm{At}}(\eta)\;,  \label{conc2} \\
&\Phi_2:= \Phi_{2}(\eta)=\Phi_{\mathrm{At}}(\eta) \circ \Phi_{\mathrm{At}%
}(\eta)\;.  \label{conc1}
\end{eqnarray}
We stress that $\Phi_{1}$ and $\Phi_{2}$ act only on one of the two modes of
the incoming twin-beam. 
The entanglement properties of the two setups can be established by applying
the criterion (\ref{eq:EBTgauss})-(\ref{eq:EBTgauss2}) to $\Phi_{1,2}$. As
already recalled, in \cite{mucritico} it was shown that $\Phi_2=\Phi_{%
\mathrm{At}}^2(\eta)$ never becomes ${\mathrm{EB}} $ for any value of the
transmissivity $\eta$. On the contrary, it can be shown that $\Phi_1$, given
by 
\begin{eqnarray}
K_{1}&=&\eta K_{\mathcal{S}_1}(r) \\
l_{1}&=&0 \\
\beta_{1}&=& \left( \frac{1-\eta }{2}\right) (\eta K_{\mathcal{S}_1}(r)^2 + %
\openone)\, ,
\end{eqnarray}
is EB if and only if 
\begin{equation}  \label{eq:tildeeta}
\eta\leq \tilde{\eta}(r)=\frac{1}{2} \left(\cosh (2 r)-\sqrt{2 \cosh (2 r)-1}%
\right) \text{csch}^2(r)
\end{equation}
or equivalently 
\begin{equation}  \label{eq:rtilda}
r \geq \tilde{r}(\eta)=\frac{1}{2} \cosh ^{-1}\left(\frac{\eta ^2+1}{(\eta
-1)^2 }\right)\,.
\end{equation}
In Fig. \ref{fig:valtilde} we report plots of $\tilde \eta$ vs.\ $r$ and $%
\tilde r$ vs.\ $\eta$ to better visualize the EB regions for the two
parameters.

It follows then, that for all values of $\eta$ and $r$ fulfilling the
condition (\ref{eq:tildeeta}) [or its equivalent version (\ref{eq:rtilda})]
the channel concatenations (\ref{conc2}) and (\ref{conc1}) provide an
instance of the identities (\ref{eq:PhiUPhiEB}) and (\ref{eq:PhiUPhiEB1}).
Consequently, following the argument~(\ref{argum}) we can conclude that the
map $\mathcal{S}_1(r)\circ \Phi_{\mathrm{At}}$ is an example of Gaussian
channel that is EB of order 2, and can be amended by the filtering map $%
\mathcal{S}_1(r)^\dagger=\mathcal{S}_1(-r)$: 
\begin{equation}
(\mathcal{S}_1(r)\circ\Phi_{\mathrm{At}})\circ \mathcal{S}_1(-r)\circ (%
\mathcal{S}_1(r)\circ\Phi_{\mathrm{At}}) = \mathcal{S}_1(r)\circ\Phi_{%
\mathrm{At}}^2 \notin {\mathrm{EB}}
\end{equation}
for all $\eta$'s.

\subsubsection{Experimental test}

\label{sec:entW} We conclude this section, by introducing an experimental
proposal for testing the entanglement-breaking properties of the maps
discussed above. A possible procedure is to use in both setups the product
criterion given in  Eq.\ (\ref{crit}) 
in order to test the entanglement
of the twin-beam after applying $\Phi_1$ and $\Phi_2$ [i.e. the entanglement
of the states $\rho_{\Phi_1}$ and $\rho_{\Phi_2}$]. Otherwise, if we are
able to measure the full covariance matrix of the state, we can apply the
optimal criterion of Eq.\ (\ref{eq:nuquadromis}). We will take into account
both criterions since the first one could be experimentally simpler
while the second one provides a conclusive answer.

In our case, the covariance matrix for $\rho_{\Phi_1}$ is given by 
\begin{equation}
\mathbf{V}_\rho=\begin{pmatrix} \alpha(\eta,r,r') & 0 &\gamma(\eta,r,r')
&0\\ 0&\alpha(\eta,-r,r')&0&-\gamma(\eta,-r,r') \\ \gamma(\eta,r,r') & 0
&\frac{1}{2}\cosh r' &0\\0&-\gamma(\eta,-r,r')&0&\frac{1}{2}\cosh r'
\end{pmatrix}\,,
\end{equation}
where 
\begin{eqnarray}
\alpha(\eta,r,r^{\prime })&=&\frac{1}{2} \left(e^{2 r} \eta \left(\eta \cosh
\left(r^{\prime }\right)-\eta+1\right)-\eta+1\right)  \nonumber \\
\gamma(\eta,r,r^{\prime })&=&-\frac{1}{2} e^r \eta \sinh \left(r^{\prime
}\right)\,.
\end{eqnarray}
If follows that $\left\langle Q ^2 \right\rangle$ and $\left \langle P
^2\right\rangle$ in (\ref{eq:wgauss}) are given by 
\begin{eqnarray}
\left\langle Q ^2 \right\rangle&=& \frac{1}{4} \left(\cosh \left(r^{\prime
}\right)+2 \alpha(\eta,r,r^{\prime }) +4 \gamma(\eta,r,r^{\prime }) \right)
\\
\left\langle P^2 \right\rangle&=&\frac{1}{4} \left(\cosh \left(r^{\prime
}\right)+2 \alpha(\eta,-r,r^{\prime }) -4 \gamma(\eta,-r,r^{\prime }) \right)
\nonumber \\
\end{eqnarray}
and for what concerns the computation of $\nu^2$ we get 
\begin{eqnarray}
\Sigma&=&\frac{\cosh ^2(R)}{4}+\alpha(\eta,r,r^{\prime })
\alpha(\eta,-r,r^{\prime })  \nonumber \\
&&+2 \gamma(\eta,r,r^{\prime }) \gamma(\eta,-r,r^{\prime }) \\
\det[\mathbf{V}]&=&-\frac{1}{4} \left(2 \gamma(\eta,r,r^{\prime
2}-\alpha(\eta,r,r^{\prime }) \cosh (R)\right)  \nonumber \\
&&\times \left(\alpha(\eta,-r,r^{\prime }) \cosh (R)-2
\gamma(\eta,-r,r^{\prime 2}\right)\,.  \nonumber
\end{eqnarray}
As already observed, the state $\rho_{\Phi_2}$ which describes the system at
the output of the second configuration can be obtained from $\rho_{\Phi_1}$
by simply setting $r=0$: therefore, in this same limit the above equations
can also be used to determine the corresponding values for the state $%
\rho_{\Phi_2}$. 

\begin{figure}[h]
\centering
\subfigure[] {\label{subfig:r1}%
\includegraphics[width=0.9\columnwidth]{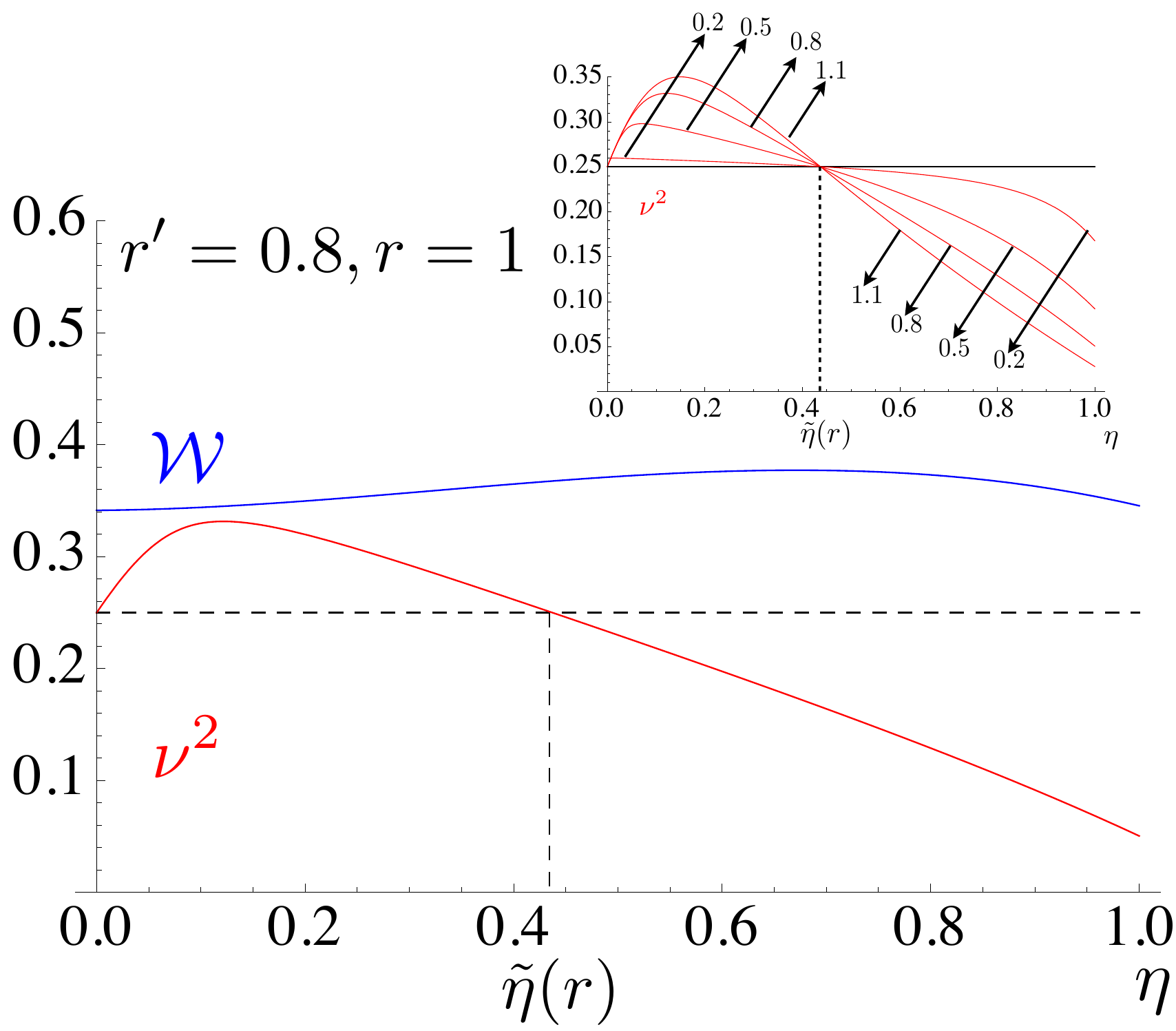}}
\hspace{15mm}
\subfigure[] {\label{subfig:r0}\includegraphics[width=0.9%
\columnwidth]{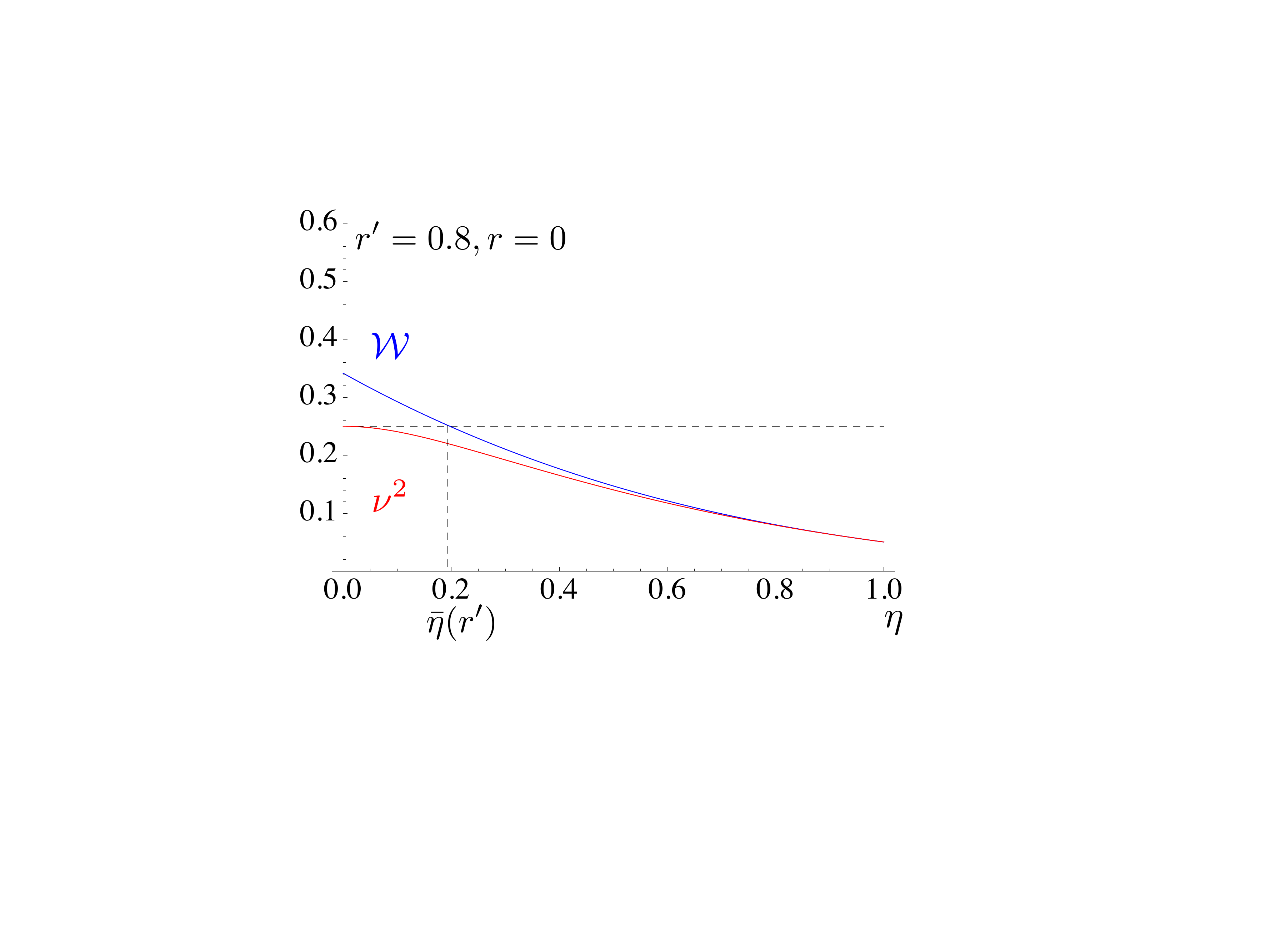} }
\caption{(Color online) 
Gaussian witness $\mathcal{W}$ (blue lines) and
theoretical test $\protect\nu^2$ (red lines) for the setup 1 of  Fig.~\ref{fig:simplifiedsetups} ($r=1$, subfigure (a)) and 
for the setup 2 of Fig.~\ref{fig:simplifiedsetups} ($r=0$,  subfigure~(b)).
In both cases the two-mode squeezing parameter of the initial state has been fixed to $r'=0.8$.
The inset of subfigure (a) shows instead the behavior of $\nu^2$ for different values of $r'$. Here, one can verify that
the EB threshold $\tilde \eta$ is independent from the initial entanglement as a consequence of the property introduced
in Sec.\ \ref{sec:Choi}. However, larger values of $r'$ allow for a clean-cut discrimination of the two regions.
}
\label{fig:r0andr1}
\end{figure}

The results for both channels 
are presented in Fig. \ref{fig:r0andr1} which shows the  values of  $\mathcal W$ and 
$\nu^2$ as functions of the beam splitter transmittivity $\eta$. The comparison with the
entanglement measure $\nu^2$ is  useful  to 
determine the values of $\eta$ and $r$ for which 
 the product criterion  provides a reliable entanglement test. 
In the second
setup [$r=0$] we expect the state of the twin-beam to be entangled, since $%
\Phi_{At}(\eta)^2\notin {\mathrm{EB}} $ for all $\eta$'s.
 On the one hand, as expected we have that $\nu^2$ is always
lower that $1/4$, the bound being saturated when $r^{\prime }=0$ or $\eta=0$
(see Fig. \ref{subfig:r0}).
On the other hand, for $%
\eta\leq \bar{\eta}$ 
\begin{equation}
\bar{\eta}(r^{\prime })=\tanh \left(\frac{r^{\prime }}{4}\right)
\end{equation}
we get $\mathcal{W}>1/4$, and thus we cannot distinguish $\rho_{\Phi_{2}}$
from a separable state if the product criterion is used.
We conclude that the product criterion, directly accessible by
a dual homodyne set-up, is reliable for $\eta \geq \bar{\eta}$. On the contrary the
PPT criterion, requiring the full experimental reconstruction of the state
covariance matrix, can be used all the way down to $\eta=0$,
as shown in Fig. \ref{subfig:r0}.

If we switch on the optical squeezer [$r>0$] for $r\geq \tilde{r}(\eta)$
(see Eq. (\ref{eq:rtilda})), we will get $\nu^2\geq1/4$ and the same we
expect for $\mathcal{W}$, as $\Phi_1 \in {\mathrm{EB}} $. Equivalently, for
any fixed $r$, from Eq.(\ref{eq:tildeeta}) we know that $\Phi_1 \in {\mathrm{%
EB}}$ for $\eta \leq \tilde{\eta}(r)$, as also proved by the behavior of $%
\nu^2$ in Fig. \ref{subfig:r1} where we have set $r=1$. On the contrary, $%
\mathcal{W}$ is always greater than $1/4$, and thus our test based on
$\mathcal{W}$ is not conclusive 
for $\eta \geq \tilde{\eta}(r)$. This comes from the fact that the product
criterion, while being directly accessible by measurements, gives a sufficient
but not necessary condition for entanglment.\newline

Summarizing if we fix the squeezing parameter $r$, in order to get a
reliable test by measuring $\mathcal{W}$ for both setups, the transmissivity 
$\eta$ of the beamsplitter should be fixed such that 
\begin{equation}  \label{eq:reliability}
\bar{\eta}(r^{\prime }) \leq \eta \leq \tilde{\eta}(r)\,.
\end{equation}
Under these conditions the witness measurement we have selected allows us to
verify that $\rho_{\Phi_2}$ is entangled [meaning that $\Phi_2$ is not EB].
At the same time the state $\rho_{\Phi_1}$ will not pass the entanglement
witness criterion in agreement with the fact that $\Phi_1$ is EB. Of course
this last result can not be used as an experimental \emph{proof} that $%
\Phi_1 $ is EB since, to do so, we should first check that no other
entanglement witness bound is violated by $\rho_{\Phi_1}$. Notice that this drawback can be avoided if we are able to
compute the optimal witness $\nu^2$ by measuring the full covariance matrix
of the output state. Finally, let us stress that $\tilde{\eta}(r)$ in the final relation (\ref%
{eq:reliability}) does not depend on the two-mode squeezing of
the incoming twin-beam (see inset of Fig. \ref{fig:r0andr1} (a) ) and thus we do not need to test the EB
properties of our maps on states characterized by an infinite amount of
energy, that is on maximally-entangled states. This represents an
important observation, especially from the point of view of the experimental
implementation of our scheme. A more detailed analysis of possible experimental
losses and detection errors will be addressed in a future work \cite{exp}.

\subsection{Example 2: asymmetric noise-phase shift-asymmetric noise}\label{EX2} 

In the previous section we have seen a class of EB Gaussian channels which
are amendable through a squeezing filtering transformation $\mathcal{S}(r)$.
Here we focus on channels which are amendable with a different unitary
filter: a phase shift $\mathcal{R}(\theta)$. According to the previous
notation, the phase shift $\mathcal{R}(\theta)$ can be represented with the
triplet: 
\begin{eqnarray}
K_{\mathcal{R}}&=&R(\theta)^T \\
l_{\mathcal{R}}&=&0 \\
\beta_{\mathcal{R}}&=&0
\end{eqnarray}
where 
\begin{equation}
R(\theta)=\begin{pmatrix} \cos(\theta) & \sin (\theta) \\ -\sin (\theta)&
\cos (\theta) \, \end{pmatrix}
\end{equation}
is a phase space rotation of an angle $\theta$.

Following the analogy with the previous case we look for a channel $\Phi$,
such that the concatenation 
\begin{equation}
\Phi \circ \mathcal{R}(\theta)\circ \Phi
\end{equation}
is EB or not EB, depending on the value of $\theta$.

It is easy to check that $\Phi$ cannot be an attenuation channel because in
this case it would simply commute with the filtering operation $\mathcal{R}%
(\theta)$. A good candidate is instead the channel $\Phi_{\mathcal{P}%
}(\eta,N_{\mathcal{P}})$, given by 
\begin{eqnarray}
K_{\mathcal{P}}&=&\sqrt{\eta} \openone \\
l_{\mathcal{P}}&=&0 \\
\beta_{\mathcal{P}}&=& N_{\mathcal{P}} \Pi+\frac{1-\eta}{2} \openone
\end{eqnarray}
where $\Pi=\left(%
\begin{array}{cc}
0 & 0 \\ 
0 & 1%
\end{array}%
\right)$, $0\le \eta \le 1$ and $N_{\mathcal{P}} \ge 0$. Notice that this
corresponds to an attenuation channel where the noise affects only the $P$
quadrature of the mode. This channel does not commute with a phase shift $%
\mathcal{R}(\theta)$ and, as we are going to show, the composition $\Phi_{%
\mathcal{PRP}}(\theta)=\Phi_{\mathcal{P}}\circ \mathcal{R}(\theta) \circ
\Phi_{\mathcal{P}}$ is EB only for some values of the angle $\theta$.

From the composition law in Eq.\ (\ref{eq:composition}) we have that the
total map $\Phi_{\mathcal{PRP}}(\theta)$ is given by 
\begin{eqnarray}
K_{\mathcal{PRP}}&=&\eta R(\theta) \\
l_{\mathcal{PRP}}&=&0 \\
\beta_{\mathcal{PRP}}&=& N_{\mathcal{P}} \left(\eta R(\theta) \Pi
R(\theta)^T + \Pi \right)+\frac{1-\eta^2}{2}\openone\,.
\end{eqnarray}
The entanglement breaking condition given in Eq.\ (\ref{eq:EBTgauss}), is
equivalent to $\nu^2\ge 1/4$ as explained in Sec.\ \ref{sec:Choi}. This
implies that

\begin{equation}
\Phi_{\mathcal{PRP}}(\theta) \text{ is EB}\Longleftrightarrow \nu^2 \ge 
\frac{1}{4} \Longleftrightarrow \theta_{min} \le \theta \le \theta_{max},
\label{EBtheta}
\end{equation}
where $\theta_{min}$ and $\theta_{max}$ are solutions of the equation $%
\nu(\theta)^2=1/4$. They can be explicitly determined: $\theta_{min}=\text{%
arcos}(\sqrt{c})$ and $\theta_{max}=\text{arcos}(-\sqrt{c})$, where 
\begin{equation}
c =\frac{2 \eta N_{\mathcal{P}}^2-2 \eta ^2-(\eta -1) (\eta +1)^2 N_{%
\mathcal{P}}}{2 \eta N_{\mathcal{P}}^2}.
\label{c-eta}
\end{equation}
The two solutions make sense only in the cases in which $0 \le c \le 1$. We may identify this as an
\textit{amendability condition}. Otherwise, in the cases in
which there are no admissible solutions, it means that the channel is
constantly EB or not EB independently of the filtering operation.

\subsubsection{Experimental test}

If we want to experimentally test the EB property of the channel $\Phi_{%
\mathcal{PRP}}(\theta)$ as a function of the filtering parameter $\theta$,
we should be able to realize the operations $\Phi_{\mathcal{P}}(\eta,N_{%
\mathcal{P}})$ and $\mathcal{R}(\theta)$.

A phase shift operation $\mathcal{R}(\theta)$ applied to an optical mode can
be realized by changing the effective optical path. This is a classical
passive operation and it is experimentally very simple. The main difficulty
is now the realization of the channel $\Phi_{\mathcal{P}}(\eta,N_{\mathcal{P}%
})$. A possible way to realize $\Phi_{\mathcal{P}}(\eta,N_{\mathcal{P}})$ is
to combine a beam splitter with an additive phase noise channel $\mathcal{N}%
(N_{\mathcal{P}})$. This is defined by the triplet 
\begin{eqnarray}
K_{\mathcal{N}}&=&0 \\
l_{\mathcal{N}}&=&0 \\
\beta_{\mathcal{N}}&=& N_{\mathcal{P}} \Pi
\end{eqnarray}
and is it essentially a random displacement $W(\delta,0)$ of the $P$
quadrature, where the shift $\delta$ is drawn from a Gaussian distribution
of variance $N_{\mathcal{P}}$ and mean equal to zero. This could be realized
via an electro-optical phase modulator driven with classical electronic
noise or by other techniques. It is immediate to check that $\Phi_{\mathcal{P%
}}(\eta,N_{\mathcal{P}})= \mathcal{N}(N_{\mathcal{P}}) \circ \Phi_{\mathrm{At%
}}(\eta,0), $ \textit{i.e.} a beam splitter followed by classical phase
noise is a possible experimental realization of the channel $\Phi_{\mathcal{P%
}}(\eta,N_{\mathcal{P}})$.

\begin{figure}[t]
\begin{center}
\includegraphics[width=1\columnwidth]{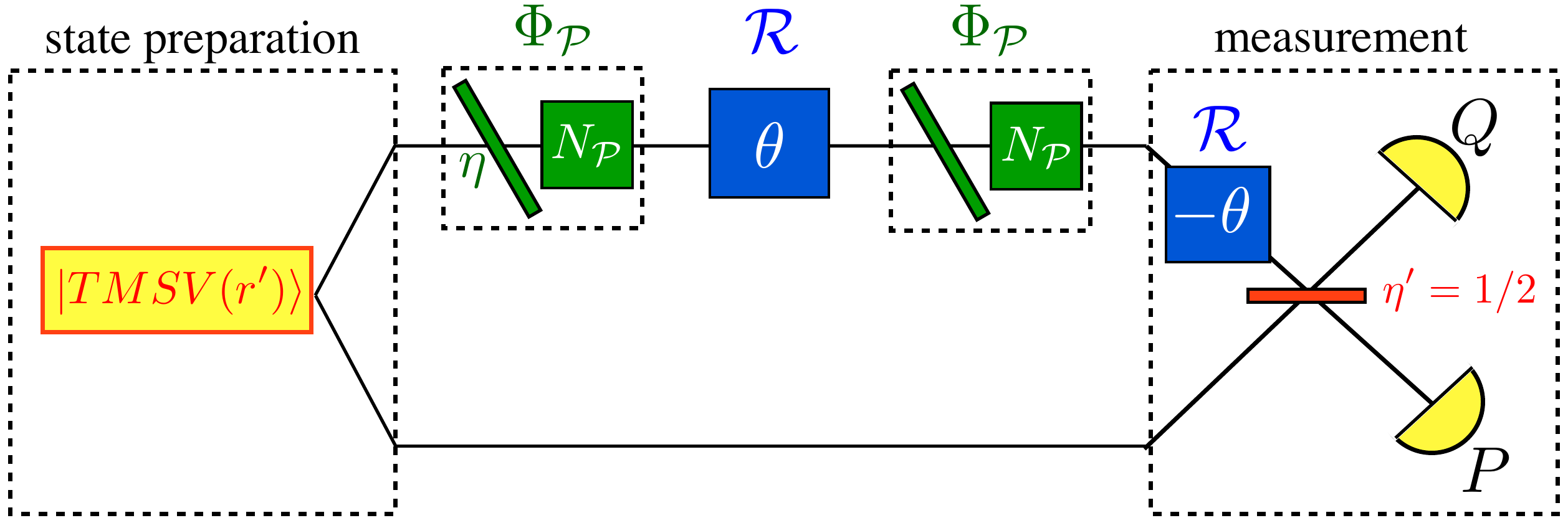}
\end{center}
\caption{
Schematic of the experimental proposal discussed in Sec.~\ref{EX2}.
As in Fig.~\ref{fig:simplifiedsetups} the setup is divided in three stages (preparation of the probing state $|TMSV\rangle$, application of the channels,
and finally measurement of the output state).   The global map is obtained by applying twice the Gaussian channel $\Phi_{\mathcal{P}}$ with the intermediate insertion  
  of a unitary phase shifter $\mathcal{R}(\theta)$. Depending on the value of the phase shift $\theta$ the global channel is EB or not. 
 }
\label{exp2}
\end{figure}

The proposed experimental setup is sketched in Fig.\ \ref{exp2}. 
A two-mode squeezed state is prepared and the desired sequence of channels
is applied on one mode of the entangled pair. The presence of entanglement
after the application of all the channels is verified by measuring the
variances of $Q$ and $P$ defined in (\ref{eq:wgauss}) after a unitary
correction $\mathcal{R}(-\theta)$. This correction does not change the
entanglement of the state but it is important for optimizing the
entanglement criterion (\ref{crit}).

A possible experiment could be to measure the witness for various choices of
the filtering operation, or in other words for various values of $\theta$.
One should check that the condition for entanglement $\mathcal{W}<1/4$ is
verified only for some angles $\theta$ while for $\theta_{min}\le \theta \le
\theta_{max}$ we must have $\mathcal{W} \ge 1/4$ because the channel is EB.
As a figure of merit for the quality of the experiment, the witness $%
\mathcal{W}$ can be compared with the corresponding optimal witness $\nu^2$.

\begin{figure}[t]
\begin{center}
\includegraphics[width=0.9\columnwidth]{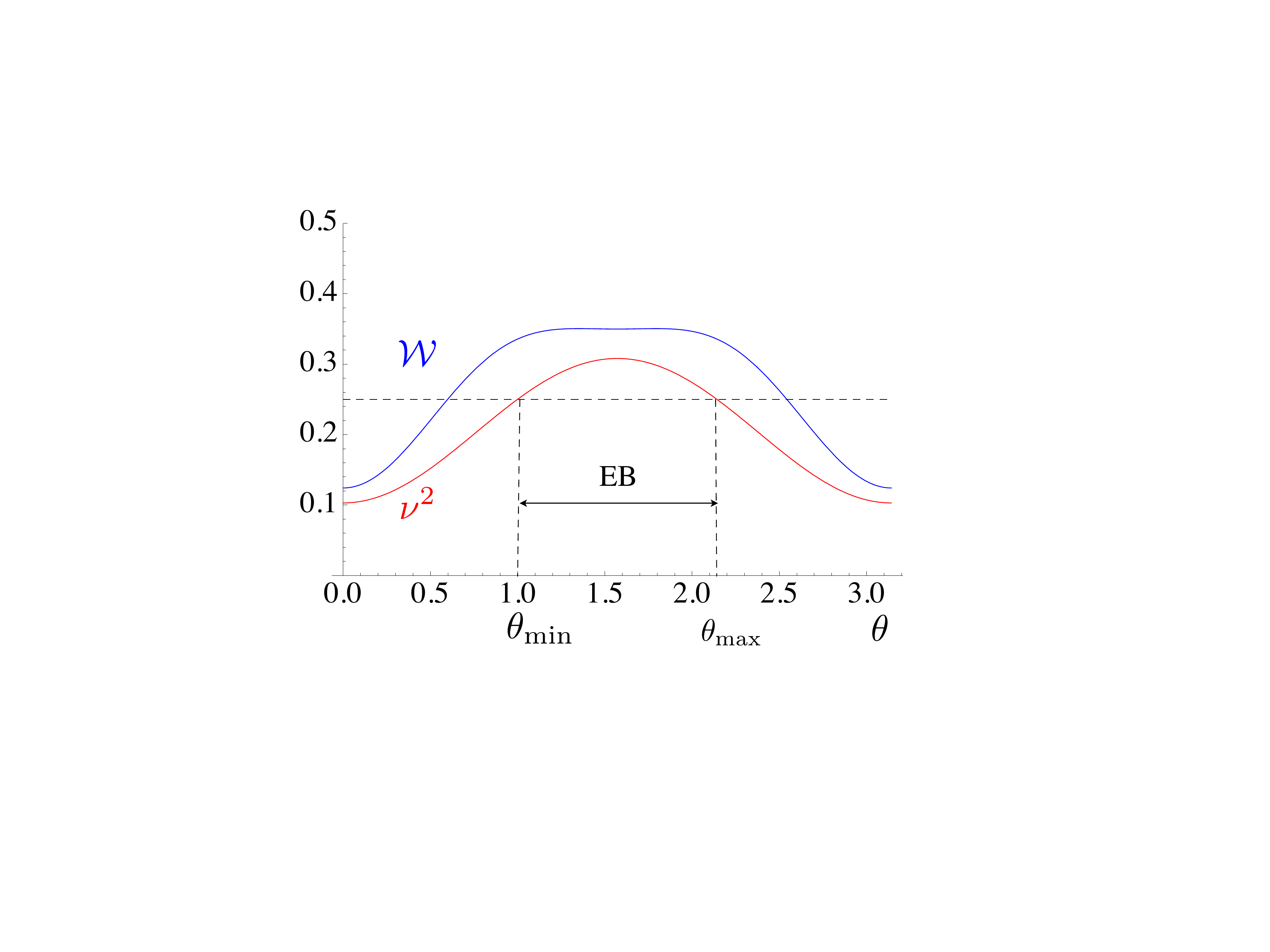}
\end{center}
\caption{Entanglement witness $\mathcal{W}$ and optimal theoretical witness $%
\protect\nu^2$ as functions of the angle $\protect\theta$ for the setup of Fig.\ \protect\ref{exp2} with parameters: $r'=2$, $%
\protect\eta=0.9$ and $N_{\mathcal{P}}=1$. In this case we find that the global
channel is entanglement breaking only in the region $\theta_{\mathrm{min}} <\theta< \theta_{\mathrm{max}}$
where, $\protect \theta_\mathrm{min}=0.99$ and $\protect\theta_{max}=2.15$.
}
\label{plot2}
\end{figure}

The results are plotted in Fig.\ \ref{plot2}. For some values of $\theta$,
one can experimentally show that the channel is not EB. On the other hand,
inside the entanglement breaking region, the witness is consistently larger
than $1/4$. Again, we underline that, if we are able to
measure the covariance matrix of the output state, the product criterion can
be replaced by the optimal one $\nu^2 < 1/4$ (see Eq.
(\ref{eq:nuquadromis})). 

As a final remark we stress that, even though
 it is realistic to consider $\eta < 1$ to account for 
 experimental losses,  
 the same qualitative results are possible
in the limit of $\eta=1$, \textit{i.e.} without the two beam splitters. In this case the amendability condition $0\le c\le1$ (see Eq. (\ref{c-eta})) implies 
$N_{\mathcal{P}}\ge 1$ and the global map is EB for
\begin{eqnarray}
\arccos\left(\sqrt{1-1/N^{2}_{\mathcal{P}}}\right)
\le \theta \le \arccos\left(-\sqrt{1-1/N^{2}_{\mathcal{P}}}\right). \nonumber 
\end{eqnarray} 

\section*{Conclusions}

In this paper we proved the existence of amendable Gaussian maps by
constructing two explicit examples. For each of them we put forward an
experimental proposal allowing the implementation of the map. We took as
benchmark model the set of entanglement breaking maps, and presented a sort of
``error correction'' technique for Gaussian channels. Differently from the
standard encoding and decoding procedures applied before and after the
action of the map \cite{nogo}, it consists in considering a composite map $%
\Phi \circ \Phi$ with $\Phi \in {\mathrm{EB}} ^2$ and applying a unitary
filter between the two actions of the channel so as to prevent the global
map from being entanglement breaking.

We focused on two-mode Gaussian systems. We recall that in order to test the
entanglement breaking properties of a map we have to apply it, tensored with
the identity, to a maximally-entangled state, which in a continuous variable
setting would require an infinite amount of energy. However in Sec.~\ref%
{sec:Choi} we have proved that without loss of generality it is sufficient
to consider a two-mode squeezed state with finite entanglement. This
property is crucial for the experimental feasibility of our schemes.
Finally, in order to verify if the entanglement of the input state survives
after the action our Gaussian maps, we applied the product criterion to the
out coming modes \cite{prodcrit}, and compared it with the
entanglement-negativity. The latter analysis enabled us to properly set the
intervals to which the experimental parameters have to belong in oder to
consider the product criterion reliable.

This analysis paves the way to a broad range of future perspectives. 
One possibility would be to extend it to the case of multimode Gaussian or non
Gaussian maps. Another compelling isuee would be determining a complete
characterization of amendable Gaussian maps of second or higher order. We
recall that, according to the definition introduced in \cite{mucritico}, a
map $\Phi$ is amendable of order $m\geq2$, if $\Phi \in {\mathrm{EB}} ^2$
and it is possible to delay its detrimental effect by $m- 2$ steps by
applying the same intermediate unitary filter after successive applications
of the channel. One possible outlook in this direction would be to allow the
choice of different filters at each error correction step and determine an
optimization procedure over the filtering maps. Of course this analysis
would be extremely difficult to be performed for arbitrary noisy maps. A
first step would be to focus on set of the Gaussian maps using the
conservation of the Gaussian character under combinations among them and
their very simple composition rules to perform this analysis.

\end{document}